# THE INVERSE-COMPTON AND EXTRAGALACTIC COMPONENTS OF THE DIFFUSE GAMMA-RAY EMISSION

A. Chen, J. Dwyer[*], and P. Kaaret

Department of Physics and Columbia Astrophysics Laboratory

Columbia University, New York, NY  10027

[*]now at Department of Physics, University of Maryland at College Park, College Park, MD 20742

## ABSTRACT

We present spectra of the inverse-Compton and extragalactic components of the high-energy gamma-radiation based on an analysis of eight high-latitude ($29°.5 < |b| < 74°$) regions. The spectrum of gamma-radiation that is correlated with atomic hydrogen (H I) column density indicates that this component originates in cosmic-ray/matter interactions. The gamma-ray emission uncorrelated with H I varies as a function of galactic longitude and is correlated with radio continuum emission (408 MHz), which is dominated by synchrotron radiation from cosmic-ray electrons. We interpret the longitude-dependent variation in the intensity of gamma-ray emission uncorrelated with H I as a model-independent, direct measure of the spatially varying part of the inverse-Compton (IC) emission from the galactic halo. We find an intensity change (center versus anticenter) of $(3.3 \pm 0.7) \times 10^{-6}$ photons cm$^{-2}$ s$^{-1}$ ster$^{-1}$ for $E > 100$ MeV, a statistically significant ($> 4\sigma$) result. This value is a lower limit to the intensity of inverse-Compton gamma-ray emission towards the galactic center at high latitudes.

We correlate the gamma-ray intensity with a model consisting of an isotropic component, a component proportional to the 408 MHz synchrotron radiation, and an HI component with different emissivities in eight galactic octants. Using this fit, we measure the spectrum and absolute intensity of the IC emission. We obtain a spectral index of $-1.85 \pm 0.17$, in agreement with measurements of the near-earth electron cosmic-ray spectrum, and an average intensity of $(5.0 \pm 0.8) \times 10^{-6}$ photons cm$^{-2}$ s$^{-1}$ ster$^{-1}$ for $E > 100$ MeV. Extrapolating the IC emission to zero galactic radio continuum intensity, we find that the extragalactic gamma-radiation has a spectral index of $-2.15 \pm 0.06$ and an intensity of $(1.24 \pm 0.06) \times 10^{-5}$ photons cm$^{-2}$ s$^{-1}$ ster$^{-1}$ for $E > 100$ MeV. The spectrum is consistent with those measured for gamma-ray loud AGN and lends support to interpretation of the extragalactic radiation as due to unresolved gamma-ray emitting AGN.

Subject headings: Gamma rays: Observations -- ISM: Cosmic Rays -- ISM: Structure -- ISM: Magnetic Fields -- Radiation Mechanisms: Miscellaneous -- Radio Continuum: Interstellar

# 1 INTRODUCTION

The gamma-ray sky above 30 MeV is dominated by diffuse emission due to the interaction between high-energy cosmic rays and interstellar matter in our Galaxy, localized along the galactic plane. Other possible sources of diffuse emission include inverse-Compton (IC) emission from scattering of cosmic-ray electrons on low-energy photons, and extragalactic radiation which may arise from a superposition of unresolved point sources. In this paper we present an observational, model-independent measurement of the inverse-Compton (IC) emission using data from EGRET (Fichtel et al. 1993). The gamma-ray observations are used to provide information about the spectrum and extent of the cosmic-ray electron halo. Measurement of the IC emission allows us to determine the spectrum and intensity of the extragalactic gamma-radiation.

Because the interstellar photon field and the cosmic-ray electron density are both difficult to observe directly, most studies of the inverse-Compton emission to date have relied on models of these quantites (Kniffen & Fichtel 1981; Bloemen 1985; Chi et al. 1989). Uncertainties in the parameters of these models, including the distribution of the optical to far infrared photon fields (Mathis et al. 1983) and the scale height of the electron halo (Chi et al. 1989 and references therein), have caused current estimates of the contribution from the inverse-Compton emission to vary widely, from less than 10% to over 50% of the total intensity for galactic latitudes greater than a few degrees. Previous attempts to measure directly the inverse-Compton component have been problematic because of the high instrumental background of COS-B (Strong et al. 1982), the brief lifetime of SAS-2 (Fichtel et al. 1975), and the fact that along the Galactic plane the emission is dominated by the cosmic-ray/matter component.

The difficulty in estimating and measuring the inverse-Compton component has resulted in uncertainty in the intensity of extragalactic emission. Studies of data from SAS-2 (Fichtel, Simpson, & Thompson 1978) established that the gamma-ray emission consists of two components: one component concentrated along the galactic plane, thought to be due primarily to cosmic-ray/matter interactions, and the other roughly isotropic, whose origin is unknown but was presumed to be extragalactic. The measured spectrum of the isotropic component was steeper than that of the only (then) known extragalactic source, 3C273 (Setti & Woltjer 1979), leading many investigators to speculate that the extragalactic emission might have a cosmologically interesting origin, rather than simply being the superposition of unresolved gamma-ray point sources. With the high-quality data available from EGRET, it is possible to separate the inverse-Compton component and accurately measure the purely isotropic emission.

In this paper, we examine the diffuse high-energy gamma-radiation in the galactic polar cap regions ($29°.5 < |b| < 74°$). At high galactic latitudes, the intensity of gamma-radiation from interactions of cosmic rays on matter is greatly reduced, facilitating the search for IC and extragalactic emission. Section 2 describes the data analysis techniques. In section 3, we examine the correlation of gamma-ray intensity with H I column density as measured by 21 cm radio data in order to separate the gamma-ray emission due to cosmic-ray interactions on matter from the IC and extragalactic emission. Section 4 presents our results on the gamma-ray emissivity at high galactic latitudes. Our results on the gamma-ray emission uncorrelated with H I are described in sections 5 and 6. In section 6, we present evidence of detection of IC emission and a new measurement of the spectrum of the extragalactic radiation. In section 7, we review our results, compare them to results from COS-B and SAS-2, and discuss the implications for models of the IC emission and the extragalactic radiation.

## 2 DATA ANALYSIS

The primary technique used in this work to distinguish diffuse emission due to cosmic-ray interactions in the Galaxy from extragalactic or inverse-Compton halo emission is linear correlation of the gamma-ray intensity versus the atomic hydrogen column density ($N_{HI}$) as measured by 21 cm radio emission. This technique was first applied by Fichtel et al. (1975) in the analysis of SAS-2 data.

The data analysis methods used here are the same as those described in Chen et al. (1995a), hereafter referred to as Paper I. The gamma-ray data were obtained from the EGRET instrument (Fichtel et al. 1993) in the form of photon counts and exposure maps with $0°.5 \times 0°.5$ pixels in different energy bins. These maps were integrated over distinct $5° \times 5°$ fields. We excluded 15° diameter circles around each of the point sources listed in the first EGRET point source catalog (Fichtel et al. 1994), totalling 22% of the sky above $|b| > 30°$. Circles of this size were required because the EGRET point spread function has non-Gaussian wings which contain a significant fraction of power and are therefore important for bright sources, particularly below 100 MeV (Thompson et al. 1993). The integrated gamma-ray data were fit to the H I column density averaged over corresponding fields using the formula $J = I + qN_{HI}$, where $J$ is the gamma ray intensity and $N_{HI}$ is the average H I column density (Figure 1). The slope, $q$, of the linear correlation is a measure of gamma-ray emissivity due to cosmic-ray interactions with atomic hydrogen. The intercept at zero column density, $I$, is an estimate of the emission not correlated with H I. This emission may have contributions from inverse-Compton scattering in the Galactic halo and extragalactic emission.

The linear correlation technique requires the assumption of uniform gamma-ray emissivity over a region of sky which shows significant variation in $N_{HI}$. Analysis of regions containing only a small range of $N_{HI}$ values produces slope and intercept estimates which are highly correlated and relatively useless for distinguishing the two components of emission. Selection of the region size must balance precision in the correlation fits versus the inaccuracy due to the approximation of constant emissivity in each region. Possible systematic flaws in the technique include errors induced in the intercept due to mass uncorrelated with H I, slope errors due to mass correlated (or anticorrelated) with H I, and errors due to spatially varying emission (e.g. inverse-Compton emission) arising from processes other than cosmic-ray interactions on H I. These sources of error are discussed in more detail below.

We have performed correlation fits for the gamma-ray intensity integrated for energies greater than 100 MeV and also in six energy bins (30-100-150-300-500-1000-$10^5$ MeV). In each case, the H I data were convolved in a manner appropriate for the EGRET point spread function in the energy band. The gamma-ray data are from EGRET phases 1 and 2 (1991 April 22 to 1993 August 24). For all analysis in this paper, we have used the composite 21 cm map of Dickey & Lockman (1990) rather than the survey of Stark et al. (1992) used in Paper I, which was restricted to the northern celestial hemisphere ($\delta > -40°$). The Dickey & Lockman data include velocities from $-250$ km s$^{-1} < v_{LSR} < 250$ km s$^{-1}$ and an opacity correction assuming $T_S = 200$ K.

After an overall systematic scale offset of (3.27±0.02) % is corrected, the two surveys are in good agreement. This offset is taken into account in comparing the results of this paper and Paper I.

The correlation fits in energy bands result in energy spectra for the emissivity (slope) and "isotropic" (intercept) emission, shown in Figure 2. We use two models to fit these energy spectra, a simple power law and a cosmic-ray induced gamma-ray emissivity model. The parameters used for our model, and typically for cosmic ray models in general, are the intensities of the electron bremsstrahlung and pion decay components, which are directly related to the densities of electron and nuclear cosmic rays. The pion decay component of the cosmic-ray model is derived from an analytic fit to a model developed by Dermer (1986), while the electron bremsstrahlung component is a power law with fixed spectral index (Bertsch et al. 1993). We note that this model produces a significantly better fit to the emissivity data than a similar model using the pion decay component from Stecker (1989). We have chosen to parameterize our model using the integrated gamma-ray flux for $E > 100$ MeV and the ratio of the electron bremsstrahlung to pion decay components ("e/p" ratio). The near-Earth cosmic-ray flux as parameterized by Dermer and Bertsch et al. results in an emissivity for $E > 100$ MeV of $1.73 \times 10^{-26}$ photons s$^{-1}$ ster$^{-1}$ H-atom$^{-1}$ and an e/p ratio defined as unity.

Lebrun et al. (1982) pointed out that galaxy counts could be used as a tracer of total mass by absorption, while 21 cm radio emission traces only H I. However, using the Lick galaxy count catalog (Seldner et al. 1977), we found that the column density as derived from galaxy counts shows a weaker correlation with the EGRET gamma-ray intensity than the 21 cm data. The gamma-ray data from EGRET have sufficient counts that the poor statistics of the galaxy counts limits the quality of the correlation. We conclude that use of galaxy counts provides no advantage in attempting to find non-H I mass.

## 3 CORRELATIONS WITH ATOMIC HYDROGEN

Our analysis of the emissivity in the galactic polar caps shows that there are significant spatial variations on large angular scales (Chen et al. 1995a, 1995b). These variations, while potentially a novel probe of galactic structure, are anathema to measurement of the extragalactic background. To allow for spatial variations in the emissivity (per H I-atom), we have chosen to divide each galactic polar cap ($29.5° < |b| < 74°$) into four offset quadrants in longitude (45°-135°-215°-315°-45°). Two sets of correlation fits were made for each quadrant: a fit of the total intensity for $E > 100$ MeV and a set of fits over six energy bands. The results of the fits are shown in Tables 1-3 and Figures 3 and 4. We note that there is good agreement between the intercepts and emissivities determined in the $E > 100$ MeV analysis and the energy-binned analysis.

The region size was selected to balance precision in the correlation fits versus the inaccuracy due to the approximation of constant emissivity in each region and the possible loss of information on spatial variations in the intercept values. In addition to adequate gamma-ray statistics, a sizable lever-arm in $N_{HI}$ across each region is required for accurate estimation of the slope and intercept. The covariance values of the $E > 100$ MeV fits are shown in Table 1. The 68% confidence contours of the error ellipses extend, on average, roughly 30% beyond the formal 1-$\sigma$ single parameter error estimates. A sizable latitude range is required to obtain an adequate lever-arm in $N_{HI}$, preferably extending to $b \sim 90°$ to sample the lowest $N_{HI}$ values. However, we used an upper limit on latitude of $|b| < 74°$ to allow a search for the inverse-Compton halo in the spatial variations of the intercepts.

The approximation of constant emissivity within each individual quadrant is reasonably good, as indicated by the $\chi^2_\nu$ values of the individual quadrant fits (see Table 1). The physical extent of each octant is only 250 pc at the scale height of H I, which is smaller than the matter/cosmic-ray coupling length expected for cosmic rays (Bertsch et al. 1993). We note that the large regions of enhanced emissivity described in Chen et al. (1995a, 1995b) tend to lie completely within the individual quadrants. Except for the quadrant ($l \sim 90°$, $b > 30°$), the $\chi^2_\nu$ for each quadrant is significantly lower than that for the complete polar cap. The large $\chi^2_\nu$ in the quadrant ($l \sim 90°$, $b > 30°$) is not surprising as the region shows strong emissivity variations.

The fraction of intensity due to cosmic-ray/matter interactions varies from 30-50% between different regions. This indicates that calculations of the intensity and, especially, the spectra of the other components of the high-latitude emission would not be accurate without first performing this correlation and removing the cosmic-ray/matter component.

## 4 EMISSIVITY

Examination of the correlation slope values in Table 1 shows a strong north-south asymmetry and significant spatial variation within the south. The hypothesis that the emissivity is uniform across both hemispheres gives a $\chi^2_\nu$ of 8.8 with seven degrees of freedom, a $7 \times 10^{-11}$ confidence level rejection. This may be partially explained by the region of enhanced emissivity in the northern hemisphere, although a similar, smaller region exists in the south (Chen et al. 1995b). The southern hemisphere value, $(1.90 \pm 0.08) \times 10^{-26}$ photons s$^{-1}$ ster$^{-1}$ H-atom$^{-1}$, is more consistent with emissivities obtained by other investigators at lower latitudes (Hunter et al. 1994 and references therein). The emissivity within the northern polar cap is consistent with the average value of $(2.98 \pm 0.10) \times 10^{-26}$ photons s$^{-1}$ ster$^{-1}$ H-atom$^{-1}$; however, there is a marginal indication that the emissivity is lower towards the anticenter. The emissivity in the southern polar cap is inconsistent with a uniform value, $\chi^2_\nu = 5.6$, $\nu = 3$. The emissivity is higher towards the galactic center.

Figure 3 shows emissivity spectra for each of the quadrants within the northern and southern galactic caps. Spectral fits to these data are shown in Table 2. We note that in all cases, except one, the cosmic-ray model produces a better fit than the power-law model. In most cases the power-law model is rejected with high confidence, while the CR model produces an adequate fit. The superior fits obtained with the cosmic-ray emissivity model support the hypothesis that the emission correlated with H I is due to cosmic ray interactions.

The ratio of the fitted electron and nucleon cosmic-ray fluxes (e/p ratio) lies in the range 0.5-1.2 in the northern hemisphere and in the range 0.4-0.7 in the southern hemisphere. The e/p ratios are consistent ($\chi^2_\nu = 0.81$) with the hypothesis of a uniform value. The mean e/p ratio, $0.63 \pm 0.08$, is significantly different from the nominal near-earth cosmic ray flux.

# 5 EMISSION UNCORRELATED WITH H I

The intercept spectra are well fit by power laws, as shown in Table 3 and Figures 2 and 4. Fits to the cosmic-ray emissivity model are generally significantly worse and are dominated by the electron bremsstrahlung component, in contradiction to the emissivity spectra. The fact that the intercepts fail to fit the cosmic-ray model in comparison to the slopes, both statistically and with respect to the e/p ratio, is consistent with the hypothesis that the emission measured by the intercepts does not contain a significant component due to cosmic-ray/matter interactions.

Emission due to interactions between cosmic rays and molecular ($H_2$) or ionized (H II) hydrogen would have the same spectrum as that due to neutral hydrogen, rather than the power law spectrum of the intercept component. We therefore conclude that $H_2$ and H II emission are not responsible for all of the uncorrelated emission, although they may contribute. Current observational evidence, including CO measurements and IRAS data, indicates a negligible average column density of molecular gas at latitudes above 30°. The model of Taylor & Cordes (1993) from pulsar dispersion measures predicts an average H II column density of $7.3 \times 10^{19}$ $H^+$ $cm^{-2}$ over the region $30° < b < 74°$. Using the emissivity calculated from the correlations in each hemisphere, H II would contribute no more than ~ 15% of the intercept. This upper limit applies only if the H II is completely uncorrelated with the H I. Any correlated component would a) not contribute to the intercept, and b) contribute to the slope, artificially inflating the emissivity; both effects would decrease the total contribution of the H II with respect to the other components.

Reynolds et al. (1995) compared velocity-resolved maps of Hα and H I in a $10° \times 12°$ region of the sky in order to determine the degree of correlation between the neutral and ionized gas. They found that although the Hα emission was only weakly correlated with the total H I column density, it was much more strongly associated with a portion of the H I (10-30%) when examined in narrow velocity intervals. They concluded that it was likely that the rest of the Hα emission was also associated with H I, although the bulk of the remaining H I would not have associated H II. Therefore, even if the total ionized and neutral hydrogen column densities were not directly correlated on large scales (of which, at high Galactic latitudes, there is little observational evidence either way), the H II would be associated with a portion of the H I. These results indicate that the fraction of the uncorrelated emission due to H II is probably less than the upper limit of 15%.

We conclude that the intercept is a good measure of the combined contribution from the extragalactic background and the IC component. The hypothesis that the intercepts for $E > 100$ MeV derive from a constant value gives a $\chi^2_\nu$ of 4.1, $\nu = 7$. This hypothesis is excluded at the $2 \times 10^{-4}$ confidence level. To search for inverse-Compton emission from the cosmic-ray electron halo of the galaxy, we have analyzed the dependence of the intercept values on galactic longitude and on the intensity of continuum radio emission.

The IC emission can be probed by examining the longitude dependence of the gamma-ray emission uncorrelated with H I. Essentially all IC models predict decreasing intensity with increasing galactocentric radius. The difference in IC intensity towards versus away from the galactic center should provide an indication of the overall IC spectrum. Figure 5 shows the correlation of intensity in four energy bands versus longitude; the resulting spectrum is shown in Figure 6. A strong correlation is present at the lower energies. We derive a spectral index of -2.2

± 0.2. Note that the magnitude of the center-anticenter difference in intensity is a direct measure of the spatially varying part of the IC emission. We find an intensity change (center versus anticenter) of $(9.7 \pm 2.7) \times 10^{-6}$ photons cm$^{-2}$ s$^{-1}$ ster$^{-1}$ in the 30-100 MeV band, and $(3.3 \pm 0.7) \times 10^{-6}$ photons cm$^{-2}$ s$^{-1}$ ster$^{-1}$ for $E > 100$ MeV, a statistically significant ($> 4\sigma$) result. These intensities are lower limits to the intensity of inverse-Compton gamma-ray emission towards the galactic center for $|b| > 30°$.

## 6 EMISSION CORRELATED WITH RADIO CONTINUUM

We have shown that there is a component of the gamma-ray emission that is uncorrelated with H I, yet shows a dependence on galactic longitude. We identified this component as inverse-Compton emission. In this section, we quantify the inverse-Compton radiation by using radio continuum synchrotron emission as a tracer of cosmic-ray electron density. The IC emission should correlate with CR electron density, much as gamma-ray emission from CR-matter interactions correlates with matter density. By simultaneously correlating the gamma-ray emission against H I and radio continuum emission, we independently determine the emission from CR-matter interaction, the inverse-Compton emission, and the isotropic background.

Inverse-Compton emission arises from the scattering of high-energy cosmic-ray electrons on low-energy interstellar photons with energies ranging from $10^{-3}$ to 10 eV. Cosmic-ray electrons in the energy range $E_e \sim$ 1-1000 GeV produce significant gamma-ray emission in the 30 MeV to 10 GeV band. The same population of electrons, moving in the galactic magnetic field, $B \sim 3$ μG, will produce synchrotron radiation with maximum intensity near the critical frequency $\nu_c = (3eB/4\pi mc)(E_e/mc^2)^2 \sim$ 48 MHz - 48 THz. The radio continuum emission from $\sim$ 100 MHz to 1 GHz is dominated by synchrotron emission. At lower frequencies the emission is subject to absorption in ionized regions, while at higher frequencies the emission is overshadowed by the 2.7 K cosmic microwave background radiation. We have chosen to use the 408 MHz all-sky survey of Haslam et al. (1982) as an indicator of the cosmic-ray electron density. We adopted an intensity of 3.7 K for the extragalactic radio emission minus the zero offset of the survey (Reich & Reich 1988) and subtracted that value from the data. This survey is most sensitive to electrons with $E_e \sim$ 5 GeV. These electrons produce $\sim$ 10-100 MeV gamma rays when scattered on photons of $\sim$ 0.1-1 eV.

### 6.1 Analysis

Unambiguous separation of the gamma-ray emission from CR-matter interactions versus IC emission is possible only if the tracers of the two emission processes are sufficiently uncorrelated. Figure 7 shows the correlation of the 408 MHz continuum radiation, $T_{408}$, versus the H I column density, $N_{HI}$. Inspection of the figure shows the two quantites are only weakly correlated, $R = 0.34$. We may therefore perform a multivariate linear regression using both quantities as independent variables.

We fit the 5° × 5° binned gamma-ray background, $J$, with a ten-parameter model,

$$J(l,b) = I + CT_{408}(l,b) + \sum_{k=1}^{8} q_k f_k(l,b) N_{HI}(l,b),$$

where $f_k(l,b) = 1$ in region $k$, and 0 otherwise. As previously, point sources were removed from the data before spatial binning and the fits were performed in the energy bands (30-100-150-300-500-1000-$10^5$ MeV) and for $E > 100$ MeV. The parameter $I$ is the isotropic background. The 8 parameters $q_k$ are the CR-matter gamma-ray emissivities in the 8 regions with $29°.5 < |b| < 74°$ used previously. The IC emissivity for CR electrons is assumed to be uniform and is given by the parameter, $C$. To test the relative merit of adding the IC radiation term, we performed an $F$-test using $F_\chi = \Delta\chi^2/\chi^2_\nu$ where $\chi^2_\nu$ is the reduced $\chi^2$ of the fit containing the extra term and $\Delta\chi^2$ is the difference between the $\chi^2$ of the fit containing the extra term and the one without. In the 30-100 MeV band, $F_\chi = 7.6$, implying a probability of 99.4% that the additional term is warranted.

In the $E > 100$ MeV energy band, the fitted parameters were $I = (1.24 \pm 0.06) \times 10^{-5}$ photons cm$^{-2}$ s$^{-1}$ ster$^{-1}$, $C = (2.23 \pm 0.4) \times 10^{-7}$ photons cm$^{-2}$ s$^{-1}$ ster$^{-1}$ K$^{-1}$, and $\chi^2_\nu = 1.49$, with $\nu = 664$. The emissivities with respect to H I, $q_k$, are shown in Table 4. The spectra of $q_k$, shown in Figure 8, are analogous to the emissivities shown in Figure 3. In all cases, the cosmic-ray model produces a better fit than the power-law model. The values obtained for the emissivity for $E > 100$ MeV, the fitted CR density, and the fitted CR e/p ratio are in reasonable agreement with the values found from the 16-parameter fit of section 2 [$\chi^2_\nu(q_k) = 2.1$, $\chi^2_\nu(CR) = 1.5$, $\chi^2_\nu(e/p) = 0.7$]. Note that one need not expect these values to be equal, since the lack of an inverse-Compton component in section 2 affects the other parameters; nevertheless, the agreement between the two fits indicates that the separation of the emission due to cosmic-ray/matter interactions is robust. The reduced covariances of the fit parameters are relatively small, indicating that the fit parameters are not highly correlated. On average, the 68% confidence contours of the error ellipsoids extend roughly 30% beyond the formal 1-$\sigma$ single parameter error estimates.

The slope, $C$, of the correlation with the 408 MHz synchrotron radiation is an indicator of the intensity of the IC emission in each band and can be used to derive an energy spectrum, as shown in Figure 9. Accepting the caveats discussed below, this spectrum can be interpreted as the energy spectrum of the IC emission.

## 6.2 Inverse-Compton Emission

The intensity of the synchrotron emission at fixed frequency varies as

$I_s \propto N_e B^{(\alpha_e+1)/2} \nu^{-(\alpha_e-1)/2} f(\alpha_e)$, where $N_e$ is the electron energy density and $f(\alpha_e)$ is a slowly varying function of the electron spectral index $\alpha_e$ (Blumenthal & Gould 1970). If $\alpha_e$ and $B$ are roughly constant in space, we can use synchrotron emission to trace the cosmic ray density (Haslam et al. 1981). The observed radio continuum spectrum can be used to constrain $\alpha_e$. Reich & Reich (1988) used 408 MHz and 1420 MHz data to obtain a map of $\alpha_e$ in the northern hemisphere. In the northern galactic polar cap $\alpha_e$ ranges from -3.0 to -2.2.

Spatial variation in the magnetic field at high $z$ is more difficult to constrain. Analysis of the local magnetic field using pulsar rotation measures yields a scale length of ~ 50 pc for the random variations in the local magnetic field (Rand & Kulkarni 1989), while the magnetic field and the cosmic-ray electron halo are both thought to extend at least 750 pc out of the plane. Therefore, although the local magnetic field varies on the order of 100%, the value of the average magnetic field in the region traversed by cosmic-ray electrons contributing to synchrotron emission along various lines of sight can be expected to vary by only ~20%.

The IC intensity depends both on the electron density and on the intensity of the interstellar radiation field. The IC intensity varies as

$$I_{IC} \propto N_e E_\gamma^{-(\alpha_e+1)/2} f_2(\alpha_e) \int \varepsilon_{ph}^{(\alpha_e-1)/2} n(\varepsilon_{ph}) d\varepsilon_{ph},$$

where $n(\varepsilon_{ph})$ is the number density of low-energy photons with energy $\varepsilon_{ph}$ (Blumenthal & Gould 1970). To show that the proportionality between the radio-synchrotron and gamma-ray IC intensities will not be strongly affected by the variation in spectral index, we first note that over the range of spectral indices of the electron spectrum, $f/f_2$ varies from 0.34 to 0.53. This implies that $\alpha_e$ depedence of the ratio of radio-synchrotron to gamma-ray IC intensity is

$$\frac{I_s}{I_{IC}}(\alpha_e) \propto \left(\frac{4\pi mc\nu}{3eB}\right)^{-(\alpha_e-1)/2} \left(\frac{E_\gamma^{(\alpha_e-1)/2}}{\left\langle \varepsilon_{ph}^{(\alpha_e-1)/2} \right\rangle}\right).$$

For 408 MHz radiation and an average magnetic field $B \sim 3~\mu G$, we find that

$4\pi mc\nu/3eB = 3 \times 10^7$. For a 1 eV photon producing a 100 MeV $\gamma$-ray, $E_\gamma/\varepsilon_{ph} \approx 10^8$. Therefore, we expect $I_S$ and $I_{IC}$ to have roughly the same dependence on $\alpha_e$.

We expect both the magnetic field and the radiation field to vary spatially. Furthermore, the degree of variation in the radiation field will depend on the energy (for example, the thermal microwave background will vary much less than the optical stellar emission). However, the average radiation field along the line of sight should vary smoothly because the scale height of the radiation field extends to greater than 1 kpc. Bloemen (1985) estimates that near the solar vicinity the energy density of even the most variable wavelength range of radiation field, the 8-1000 µm dust emission, should vary by only 10% over the range of galactrocentric radius spanned by the northern polar cap.

We find that the IC emission has a spectral index $\alpha_e = -1.88 \pm 0.14$. The intensity can be obtained by multiplying the derived IC emissivity by 22.5 K, the average synchrotron radiation brightness temperature for $|b| > 30°$ with the extragalactic component removed. We find an average intensity of $(7.9 \pm 2.9) \times 10^{-6}$ photons cm$^{-2}$ s$^{-1}$ ster$^{-1}$ in the 30-100 MeV band, and an average intensity of $(5.0 \pm 0.8) \times 10^{-6}$ photons cm$^{-2}$ s$^{-1}$ ster$^{-1}$ for $E > 100$ MeV. The spectral index of the IC emission is related to the cosmic-ray electron spectral index, $\alpha_\gamma = (\alpha_e-1)/2$. Measurements of the local cosmic-ray electron spectrum, after being corrected for the effects of solar modulation, indicate that the interstellar spectrum has an index $\alpha_e = -2.3 \pm 0.1$ at an electron energy of 1 GeV (Webber 1983), the energy which corresponds to 100 MeV IC gamma-ray emission. The implied IC emission spectral index, $\alpha_\gamma = -1.65 \pm 0.05$, is consistent with our results, especially given the uncertainty of the solar modulation corrections.

### 6.3 Extragalactic Emission

The isotropic component can be ascribed to extragalactic emission. The spectrum of the isotropic component is shown in Figure 10. We find that the extragalactic radiation has an intensity of $(3.6 \pm 0.2) \times 10^{-5}$ photons cm$^{-2}$ s$^{-1}$ ster$^{-1}$ in the 30-100 MeV band, an intensity of $(1.24 \pm 0.06) \times 10^{-5}$ photons cm$^{-2}$ s$^{-1}$ ster$^{-1}$ for $E > 100$ MeV, and a spectral index of $-2.15 \pm 0.06$.

## 7 DISCUSSION

Through examination of the spectra of the component of the gamma-ray emission correlated with H I, we have confirmed the cosmic-ray origin of the gamma-ray emissivity of interstellar matter. The quadrants closest to the galactic center have higher emissivities. A diffuse feature of enhanced emissivity toward the galactic center in the south (Chen et al. 1995b) may contribute to this effect, although the lack of contribution from the similar feature in the north (Paper I) casts doubt on that possibility. This effect may mean that cosmic ray density decreases as a function of galactic radius, or may indicate that the known concentration of molecular hydrogen toward the inner galaxy persists at higher latitudes (i.e. high $z$).

The component of the emission uncorrelated with H I was further subdivided into a longitude-dependent component that is well-correlated with synchrotron emission (a tracer of cosmic-ray electron density), which we identified as the inverse-Compton emission, and an isotropic residual, which we identified as the true extragalactic emission. Various authors, including Kniffen & Fichtel (1981), Bloemen (1985), and Chi et al. (1989), have produced models of the inverse-Compton emission by estimating the interstellar photon field as a function of position in the galaxy at various wavelengths and convolving it with a simple model of the cosmic-ray electron distribution. Differences in the estimated intensities arise primarily from different estimates of the cosmic-ray electron scale height. Bloemen found intensities of $\sim 2 \times 10^{-6}$ photons cm$^{-2}$ s$^{-1}$ ster$^{-1}$ for both the $E > 100$ MeV and 30 MeV $< E <$ 100 MeV energy bands and a spectral index of $\sim -2.0$. Kniffen & Fichtel obtained intensities of $\sim 2 \times 10^{-5}$ photons cm$^{-2}$ s$^{-1}$ ster$^{-1}$, higher than the Bloemen values because they estimated the energy density of the far-infrared photon field to be 2-4 times higher, and a spectral index of $\sim -1.8$. Chi et al. calculated intensities of $\sim 8 \times 10^{-6}$ photons cm$^{-2}$ s$^{-1}$ ster$^{-1}$ and an index of $\sim -1.95$; they used the same photon field as Bloemen (Mathis et al. 1983), but estimated the cosmic-ray electron scale height to be on the order of 5 kpc rather than 750 pc. Giller et al. (1995) estimated the IC contribution using spectral changes and emissivity variations in intermediate-latitude EGRET data and produced values similar to the Chi et al. model.

Using the longitude dependence correlations, we derived a spectral index of $-2.2 \pm 0.2$, consistent with the model predictions, and lower limits of $(9.7 \pm 2.7) \times 10^{-6}$ photons cm$^{-2}$ s$^{-1}$ ster$^{-1}$ in the 30-100 MeV band and $(3.3 \pm 0.7) \times 10^{-6}$ photons cm$^{-2}$ s$^{-1}$ ster$^{-1}$ for $E > 100$ MeV toward the galactic center. Using a model containing an isotropic component, an IC component proportional to the 408 MHz synchrotron radiation, and a spatially varying emissivity with respect to H I, we found a spectral index of $-1.88 \pm 0.14$ and average intensities of $(7.9 \pm 2.9) \times 10^{-6}$ photons cm$^{-2}$ s$^{-1}$ ster$^{-1}$ in the 30-100 MeV band and $(5.0 \pm 0.8) \times 10^{-6}$ photons cm$^{-2}$ s$^{-1}$ ster$^{-1}$ for $E > 100$ MeV. Our intensity values are higher than those of Bloemen, significantly lower than the estimate of Kniffen

& Fichtel, and marginally lower than the value of Chi et al. Assuming the photon field of Mathis et al. (1983), our value for the IC intensity is consistent with an electron scale height in the range of 1-2 kpc. The IC intensity could also be produced with an electron scale height near 750 pc and a more intense photon field.

Fichtel et al. (1978) used data from SAS-2 and 21 cm radio data to obtain a spectrum of the gamma-ray emission uncorrelated with H I, analogous to the spectrum we obtained from the simple two-parameter correlation with no IC component. However, because their statistics were insufficient to discern a center-anticenter anisotropy, they were unable to separate the uncorrelated emission into inverse-Compton and extragalactic components. Thompson & Fichtel (1982) repeated the analysis using gas column densities derived from galaxy counts instead of 21 cm data. The spectra we derived for the component of the emission uncorrelated with respect to H I may be compared directly to these SAS-2 results. We find an average spectral index of $-2.06 \pm 0.01$ and average intensities of $(1.57 \pm 0.02) \times 10^{-5}$ photons cm$^{-2}$ s$^{-1}$ ster$^{-1}$ for $E > 100$ MeV and $(4.09 \pm 0.08) \times 10^{-5}$ photons cm$^{-2}$ s$^{-1}$ ster$^{-1}$ for $30$ MeV $< E < 100$ MeV. The spectral index derived here is significantly harder than the index of $-2.7$ $(+0.4, -0.3)$ obtained by Fichtel et al., but in agreement with the index of $-2.35$ $(+0.4, -0.3)$ found by Thompson & Fichtel. The average intensities in all cases agrees with our results [$(1.0 \pm 0.5) \times 10^{-5}$ and $(1.3 \pm 0.3) \times 10^{-5}$ photons cm$^{-2}$ s$^{-1}$ ster$^{-1}$ for $E > 100$ MeV, $(4.7 \pm 1.4) \times 10^{-5}$ and $(4.2 \pm 1.4) \times 10^{-5}$ photons cm$^{-2}$ s$^{-1}$ ster$^{-1}$ for $30$ MeV $< E < 100$ MeV, respectively].

Our best estimate of the extragalactic spectrum is derived from the isotropic component of the ten-parameter fit. We find a spectral index of $-2.15 \pm 0.06$ and an intensity lower than the value for the uncorrelated component of the two-parameter fit versus H I [$(1.24 \pm 0.06) \times 10^{-5}$ photons cm$^{-2}$ s$^{-1}$ ster$^{-1}$ for $E > 100$ MeV, $(3.6 \pm 0.2) \times 10^{-5}$ photons cm$^{-2}$ s$^{-1}$ ster$^{-1}$ for $30$ MeV $< E < 100$ MeV]. Osborne et al. (1994) calculated the extragalactic spectrum by correlating high-latitude EGRET data with the gas column density as derived from 21 cm H I, CO-line H$_2$, and H$\alpha$ ionized hydrogen data, then subtracting off an inverse-Compton component derived from comparison of the model of Chi et al. (1989) with the longitude dependence of the uncorrelated residual. Only one point source was removed in the analysis. They found a spectral index of $-2.11 \pm 0.05$ and intensities of $(1.11 \pm 0.05) \times 10^{-5}$ photons cm$^{-2}$ s$^{-1}$ ster$^{-1}$ for $E > 100$ MeV and $(3.6 \pm 0.1) \times 10^{-5}$ photons cm$^{-2}$ s$^{-1}$ ster$^{-1}$ for $30$ MeV $< E < 100$ MeV. Their values for the extragalactic intensity for $30 < E < 100$ MeV and for the spectral index are in good agreement with ours, while their value for the intensity above 100 MeV is lower. Osborne et al. find a galactic north versus south asymmetry which they attribute to the extragalactic emission. We find that when the emission uncorrelated with H I is correlated with the radio synchrotron emission, no residual north/south asymmetry remains. Therefore, we attribute the asymmetry to enhanced IC emission in the northern galactic hemisphere.

A large quantity of theoretical speculation has been based on the steep spectral index for the extragalactic gamma ray background derived from SAS-2. In particular, the steepness of the spectrum has been difficult to reconcile with the simplest explanation for the origin of the extragalactic background as the sum of gamma-ray emission from blazars (Dermer & Schlickeiser 1992; Padovani et al. 1993; Stecker, Salamon, & Malkan 1993; Salamon & Stecker 1994), a class of AGN for which gamma-ray emission has been observed by EGRET in 33 cases (Fichtel et al. 1994). The need to invoke AGN evolution in order to reconcile the observed gamma ray spectra of blazars, whose mean spectral index is -2.1 ± 0.3 (Chiang et al. 1995), with the presumed spectral index of the extragalactic background has perpetuated investigation into alternative gamma-radiation mechanisms (Murthy & Wolfendale 1993 and references therein), including cosmic-ray interactions with intra-cluster gas (Dar & Shaviv 1995), inverse Compton scattering of cosmic microwave background photons from intergalactic cosmic ray electrons, primordial matter-antimatter annihilation (Gao et al. 1990), decay of Big Bang particle relics, annihilation of supersymmetric particles (Kamionkowski 1995; Stecker & Tylka 1989), and cosmic strings (Vilenkin 1988). Our analysis shows that the extragalactic gamma-ray background spectrum is consistent with typical AGN spectra. This lends support to interpretation of the extragalactic radiation as due to unresolved gamma-ray emitting AGN. Additional work in constraining the persistent emission of AGN and the AGN gamma-ray luminosity function is required to decisively test this interpretation.

We thank the EGRET instrument team. Special thanks to J. Mattox, D. Macomb, D. Kniffen, and T. McGlynn for assistance with the EGRET data. We also thank J. Condon of NRAO for supplying the H I and 408 MHz all-sky maps. We thank the referee, E. L. Wright, for useful comments. This research was supported by grant NAG 5-2235 from the NASA Compton Observatory guest investigator program.

Table 1

Correlation of gamma-ray intensity versus H I column density

| $E > 100$ MeV | | | | | |
|---|---|---|---|---|---|
| Region | Intercept | Slope | Covariance | $\chi^2_\nu$ | $\nu$ |
| N | 1.50 ± 0.03 | 3.44 ± 0.09 | -0.002 | 1.91 | 337 |
| S | 1.64 ± 0.03 | 1.90 ± 0.08 | -0.002 | 2.17 | 334 |
| $-45° < l < 45°$, N | 1.76 ± 0.07 | 3.03 ± 0.15 | -0.009 | 1.19 | 85 |
| $45° < l < 135°$, N | 1.61 ± 0.06 | 3.03 ± 0.26 | -0.015 | 2.15 | 94 |
| $135° < l < 225°$, N | 1.50 ± 0.06 | 2.50 ± 0.27 | -0.013 | 1.57 | 81 |
| $225° < l < 315°$, N | 1.74 ± 0.10 | 3.16 ± 0.23 | -0.02 | 1.73 | 90 |
| $-45° < l < 45°$, S | 1.73 ± 0.08 | 3.01 ± 0.24 | -0.018 | 1.41 | 94 |
| $45° < l < 135°$, S | 1.46 ± 0.17 | 2.09 ± 0.38 | -0.064 | 0.99 | 85 |
| $135° < l < 225°$, S | 1.37 ± 0.06 | 2.11 ± 0.10 | -0.005 | 1.69 | 87 |
| $225° < l < 315°$, S | 1.51 ± 0.06 | 1.84 ± 0.16 | -0.009 | 1.69 | 81 |

    Best-fit parameters from correlations between EGRET gamma-ray intensity and H I column density for northern and southern galactic caps and quadrants therein. N indicates $29°.5 < b < 74°$ and S indicates $-74° < b < -29°.5$. Intercepts are in units of $10^{-5}$ cm$^{-2}$ s$^{-1}$ ster$^{-1}$; slopes are in units of $10^{-26}$ s$^{-1}$ ster$^{-1}$ H-atom$^{-1}$. Covariances are in units of $10^{-31}$ cm$^{-2}$ s$^{-2}$ ster$^{-2}$ H-atom$^{-1}$.

Table 2

Spectral fits to correlation slopes

| Region | Cosmic-ray model | | | Power law model | | |
|---|---|---|---|---|---|---|
| | Density | E/p ratio | $\chi^2_\nu$ | Flux | Index | $\chi^2_\nu$ |
| N | $1.79 \pm 0.10$ | $1.07 \pm 0.12$ | 2.06 | $3.12 \pm 0.56$ | $-1.85 \pm 0.02$ | 5.23 |
| S | $1.30 \pm 0.09$ | $0.51 \pm 0.12$ | 0.67 | $1.79 \pm 0.49$ | $-1.77 \pm 0.03$ | 7.15 |
| $-45° < l < 45°$, N | $1.94 \pm 0.17$ | $0.55 \pm 0.14$ | 0.52 | $2.66 \pm 0.93$ | $-1.79 \pm 0.04$ | 5.73 |
| $45° < l < 135°$, N | $1.44 \pm 0.29$ | $1.09 \pm 0.43$ | 1.67 | $2.54 \pm 1.66$ | $-1.86 \pm 0.08$ | 1.75 |
| $135° < l < 225°$, N | $1.17 \pm 0.30$ | $1.04 \pm 0.53$ | 1.06 | $2.13 \pm 1.65$ | $-1.77 \pm 0.09$ | 0.25 |
| $225° < l < 315°$, N | $1.72 \pm 0.25$ | $1.15 \pm 0.32$ | 0.4 | $3.11 \pm 1.43$ | $-1.85 \pm 0.06$ | 0.83 |
| $-45° < l < 45°$, S | $2.12 \pm 0.27$ | $0.45 \pm 0.21$ | 0.63 | $2.87 \pm 1.46$ | $-1.72 \pm 0.06$ | 1.86 |
| $45° < l < 135°$, S | $1.57 \pm 0.43$ | $0.43 \pm 0.42$ | 0.21 | $2.08 \pm 2.28$ | $-1.73 \pm 0.13$ | 0.43 |
| $135° < l < 225°$, S | $1.37 \pm 0.12$ | $0.62 \pm 0.16$ | 0.44 | $1.97 \pm 0.67$ | $-1.80 \pm 0.04$ | 4.88 |
| $225° < l < 315°$, S | $1.27 \pm 0.19$ | $0.64 \pm 0.28$ | 1.25 | $1.90 \pm 1.04$ | $-1.76 \pm 0.07$ | 1.62 |

Best-fit parameters of the differential energy spectra of slopes derived from correlations between EGRET gamma-ray intensity and H I column density. The correlations were performed in the northern and southern galactic caps, and in quadrants therein. Two models were fit to the spectra: a cosmic-ray/matter model and a simple power law. Density represents the ratio of the cosmic-ray proton density to the local value. E/p ratio represents the ratio between the cosmic-ray electron and proton densities. Power law flux is determined for $E > 100$ MeV and has units of $10^{-26}$ s$^{-1}$ ster$^{-1}$ H-atom$^{-1}$. N indicates $29°.5 < b < 74°$ and S indicates $-74° < b < -29°.5$. $\nu = 4$ for each of the fits.

Table 3

Spectral fits to correlation intercepts

| Region | Cosmic ray model | | | Power law model | | |
| --- | --- | --- | --- | --- | --- | --- |
| | Density | E/p ratio | $\chi^2_\nu$ | Flux | Index | $\chi^2_\nu$ |
| N | 1.05 ± 0.07 | 3.34 ± 0.28 | 7.9 | 1.53 ± 0.19 | -2.04 ± 0.02 | 1.04 |
| S | 0.92 ± 0.09 | 4.38 ± 0.50 | 5.6 | 1.59 ± 0.24 | -2.09 ± 0.02 | 1.46 |
| $-45° < l < 45°$, N | 0.97 ± 0.17 | 4.50 ± 0.92 | 2.8 | 1.73 ± 0.44 | -2.09 ± 0.03 | 1.08 |
| $45° < l < 135°$, N | 1.28 ± 0.15 | 2.87 ± 0.45 | 2.5 | 1.69 ± 0.41 | -2.01 ± 0.03 | 1.1 |
| $135° < l < 225°$, N | 1.17 ± 0.14 | 2.93 ± 0.46 | 0.5 | 1.54 ± 0.37 | -2.04 ± 0.03 | 0.21 |
| $225° < l < 315°$, N | 1.11 ± 0.23 | 3.56 ± 0.93 | 2.3 | 1.68 ± 0.62 | -2.06 ± 0.05 | 1.2 |
| $-45° < l < 45°$, S | 0.88 ± 0.21 | 5.00 ± 1.41 | 2.2 | 1.68 ± 0.57 | -2.11 ± 0.05 | 0.98 |
| $45° < l < 135°$, S | 0.49 ± 0.43 | 8.08 ± 7.98 | 0.15 | 1.31 ± 1.13 | -2.21 ± 0.12 | 0.16 |
| $135° < l < 225°$, S | 0.66 ± 0.16 | 5.23 ± 1.51 | 1.6 | 1.30 ± 0.43 | -2.12 ± 0.05 | 0.93 |
| $225° < l < 315°$, S | 0.75 ± 0.16 | 5.01 ± 1.25 | 2.1 | 1.43 ± 0.43 | -2.12 ± 0.09 | 1.16 |

Best-fit parameters of the differential energy spectra of intercepts derived from correlations between EGRET gamma-ray intensity and H I column density. The correlations were performed in the northern and southern galactic caps, and in quadrants therein. Two models were fit to the spectra: a cosmic-ray/matter model and a simple power law. Density is in arbitrary units, included for comparison of relative magnitudes only. E/p ratio represents the ratio between the cosmic-ray electron and proton densities, normalized to the local ratio. Power law flux is determined for $E > 100$ MeV and has units of $10^{-5}$ cm$^{-2}$ s$^{-1}$ ster$^{-1}$. N indicates $29°.5 < b < 74°$ and S indicates $-74° < b < -29°.5$. $\nu = 4$ for each of the fits.

Table 4

Values of $q_k$ for $E > 100$ MeV and spectral fits

| Region | $E > 100$ MeV $q_k$ | Cosmic ray model | | | Power law |
|---|---|---|---|---|---|
| | | Density | E/p ratio | $\chi^2_\nu$ | $\chi^2_\nu$ |
| $-45° < l < 45°$, N | $2.85 \pm 0.13$ | $1.63 \pm 0.14$ | $0.84 \pm 0.17$ | 0.84 | 3.06 |
| $45° < l < 135°$, N | $2.90 \pm 0.16$ | $1.78 \pm 0.17$ | $0.73 \pm 0.18$ | 1.34 | 2.23 |
| $135° < l < 225°$, N | $2.13 \pm 0.17$ | $1.48 \pm 0.19$ | $0.35 \pm 0.19$ | 0.44 | 1.78 |
| $225° < l < 315°$, N | $3.43 \pm 0.10$ | $1.84 \pm 0.11$ | $1.11 \pm 0.13$ | 2.41 | 3.85 |
| $-45° < l < 45°$, S | $3.13 \pm 0.14$ | $1.91 \pm 0.15$ | $0.75 \pm 0.14$ | 1.41 | 2.17 |
| $45° < l < 135°$, S | $1.66 \pm 0.11$ | $0.97 \pm 0.12$ | $0.74 \pm 0.23$ | 0.59 | 1.41 |
| $135° < l < 225°$, S | $1.75 \pm 0.07$ | $1.10 \pm 0.08$ | $0.65 \pm 0.12$ | 0.54 | 6.09 |
| $225° < l < 315°$, S | $1.51 \pm 0.10$ | $0.93 \pm 0.11$ | $0.78 \pm 0.23$ | 0.48 | 1.40 |

Emissivities with respect to H I column density, $q_k$, derived from a ten-parameter fit of the EGRET gamma-ray intensity to a model consisting of an isotropic component, a component proportional to 408 MHz continuum radiation, and a component with different emissivities in each of the eight quadrants of the northern and southern caps. The emissivities for $E > 100$ MeV have units of $10^{-26}$ s$^{-1}$ ster$^{-1}$ H-atom$^{-1}$. Two models were fit to the differential energy spectra obtained from the emissivities in energy bands: a cosmic-ray/matter model and a simple power law. Density represents the ratio of the cosmic-ray proton density to the local value. E/p ratio represents the ratio between the cosmic-ray electron and proton densities. $\nu = 4$ for each of the fits.

## Figure Legends

Fig 1. Linear correlation between EGRET gamma-ray intensity for energies $E > 100$ MeV, $J$, and the atomic hydrogen column density, $N_{HI}$, for the northern Galactic hemisphere ($b > 30°$). Each point represents an integration of gamma-ray intensity and H I column density over a distinct field approximately $5° \times 5°$ in size. The straight line is a least-sqares linear fit to the data. This figure first appeared in Chen et al. (1995a).

Fig 2. Differential energy spectra of the slopes and intercepts of the linear correlation between EGRET gamma-ray intensity and H I column density for the northern and southern galactic caps. The solid lines represent the best-fit power law spectra. The dashed lines are fits to a cosmic-ray/matter model with an analytic approximation to the pion-decay model of Dermer (1986) and a power-law bremsstrahlung component (Bertsch et al. 1993). The spectra have been multiplied by $E^2$ to aid the eye. The parameters of the fits are listed in Tables 2 and 3.

Fig 3. Differential energy spectra of the slopes of the linear correlation between EGRET gamma-ray intensity and H I column density in the quadrants of the northern and southern galactic caps. The fits are the same fits listed in Fig. 2. The parameters of the fits are listed in Table 2.

Fig 4. Differential energy spectra of the intercepts of the linear correlation between EGRET gamma-ray intensity and H I column density in the quadrants of the northern and southern galactic caps. The fits are the same fits listed in Fig. 2. The parameters of the fits are listed in Table 3.

Fig 5. Linear correlations between the intensity of gamma-ray emission uncorrelated with H I (the "intercept", $I$) and the average of $\cos(l)$, where $l$ is the galactic longitude, in four energy bands. The diamonds represent quadrants in the northern hemisphere, the triangles quadrants in the southern hemisphere. The straight lines are least-squares linear fits to the data.

Fig 6. Differential energy spectrum of the inverse-Compton emission as determined from the slopes of correlations between the intensity of gamma-ray emission uncorrelated with H I and the average of $\cos(l)$, where $l$ is the galactic longitude. The solid line represents the best-fit power law. The power law has a spectral index of $-2.21 \pm 0.25$.

Fig 7. Linear correlation between radio continuum radiation at 408 MHz, $T_{408}$, and the atomic hydrogen column density, $N_{HI}$, for $|b| < 74°$). Each point represents the average antenna temperature and H I column density over a distinct field approximately $5° \times 5°$ in size. The straight line is a least-sqares linear fit to the data. The linear correlation coefficient $R = 0.34$.

Fig 8. Differential energy spectra of the slopes of the linear correlation between EGRET gamma-ray intensity and H I column density in the quadrants of the northern and southern galactic

caps, as determined in a ten-parameter fit with an isotropic component and a continuum radiation component. The parameters of the fits are listed in Table 4.

Fig 9. Differential energy spectrum of the inverse-Compton emission as determined from the slopes of the linear correlation between EGRET gamma ray intensity and continuum radiation at 408 MHz as determined in a ten-parameter fit with the emissivity with respect to HI in the quadrants of the galactic caps. The solid line represents the best-fit power law. The power law has a spectral index of -1.88 ± 0.14.

Fig 10. Differential energy spectrum of the extragalactic emission as determined from the intercepts of the ten-parameter linear correlation between EGRET gamma-ray intensity, continuum radiation at 408 MHZ, and H I column density in the quadrants of the galactic caps. The solid line represents the best-fit power law. The power law has a spectral index of -2.15 ± 0.06 and an integrated flux above 100 MeV of $(1.24 \pm 0.06) \times 10^{-5}$ photons cm$^{-2}$ s$^{-1}$ ster$^{-1}$.

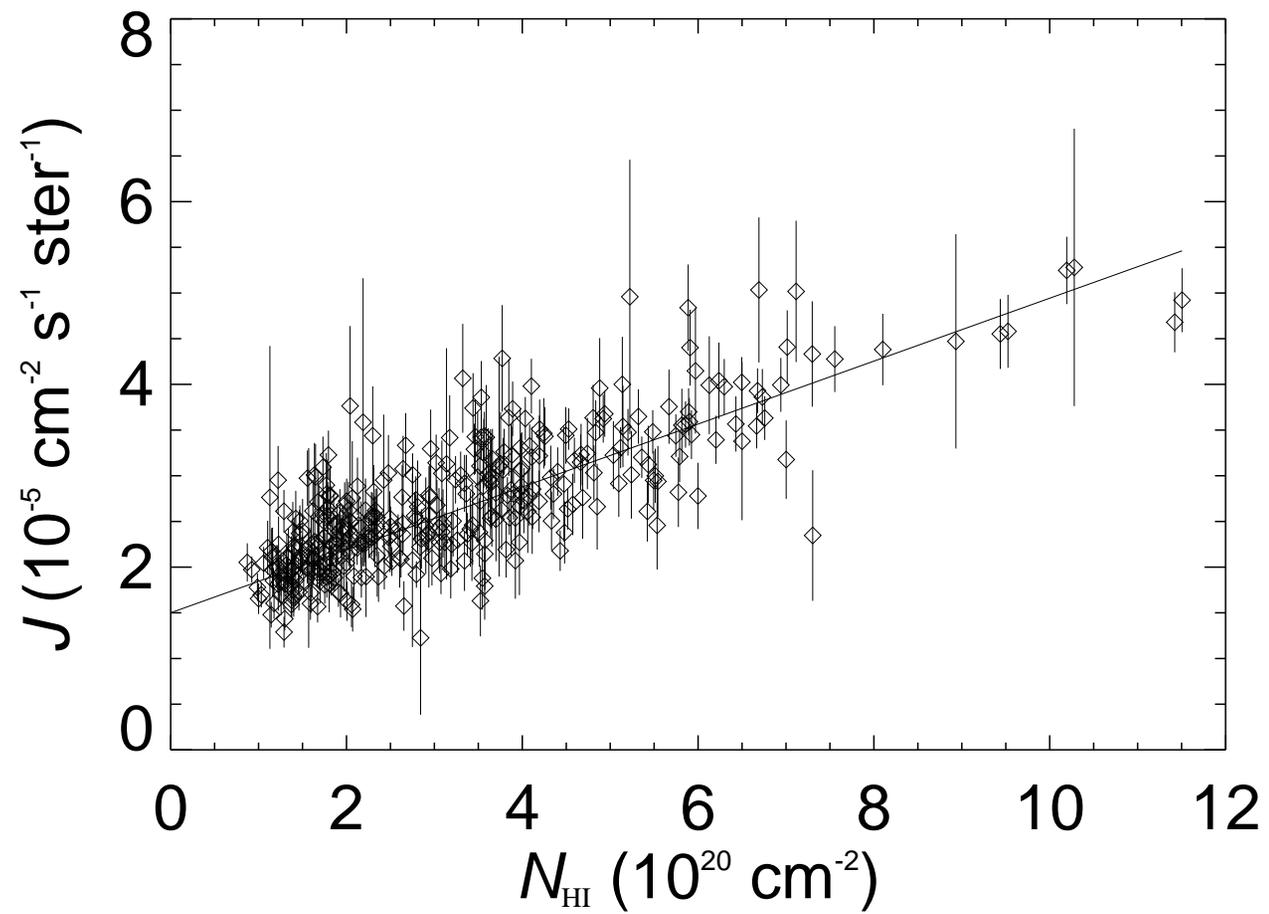

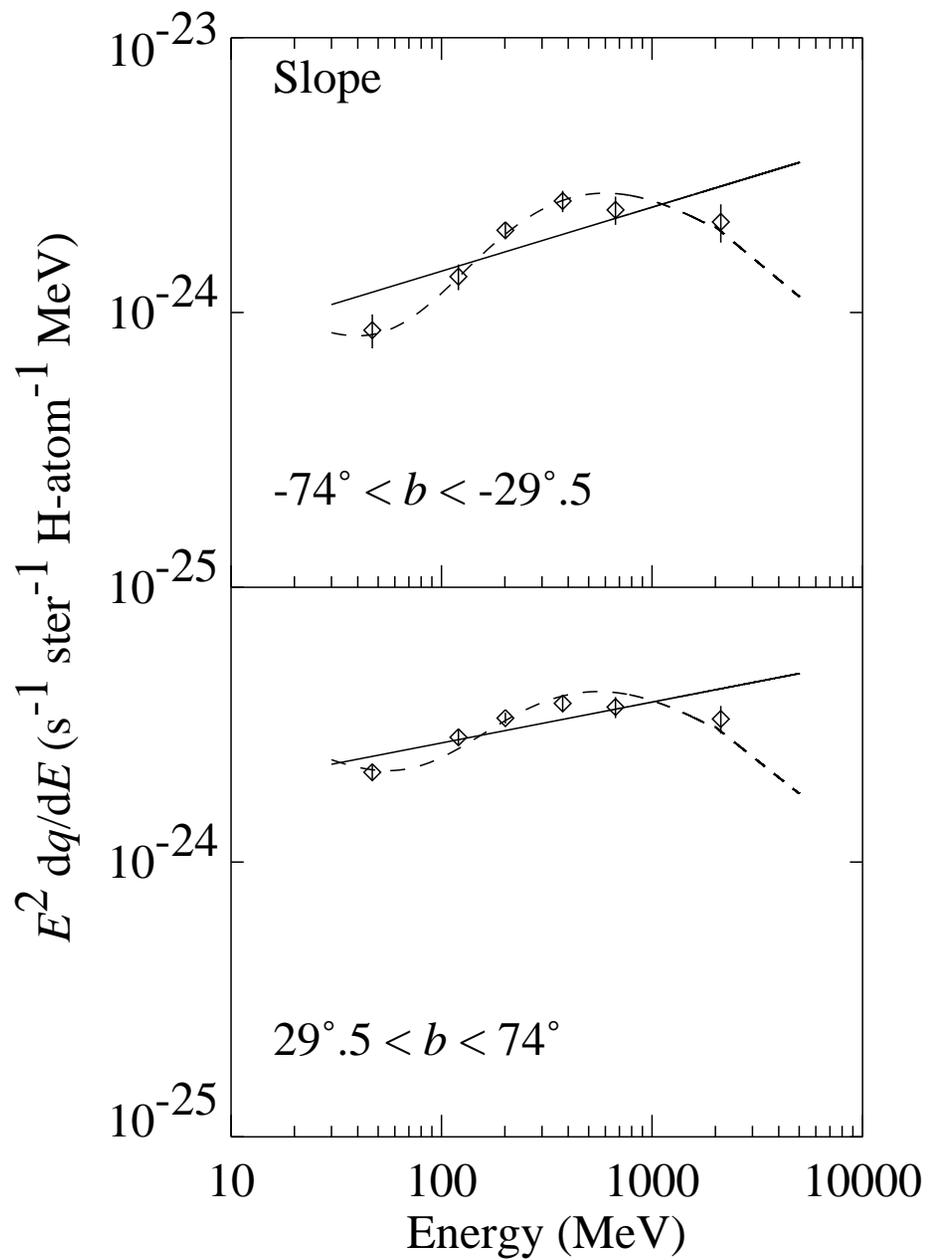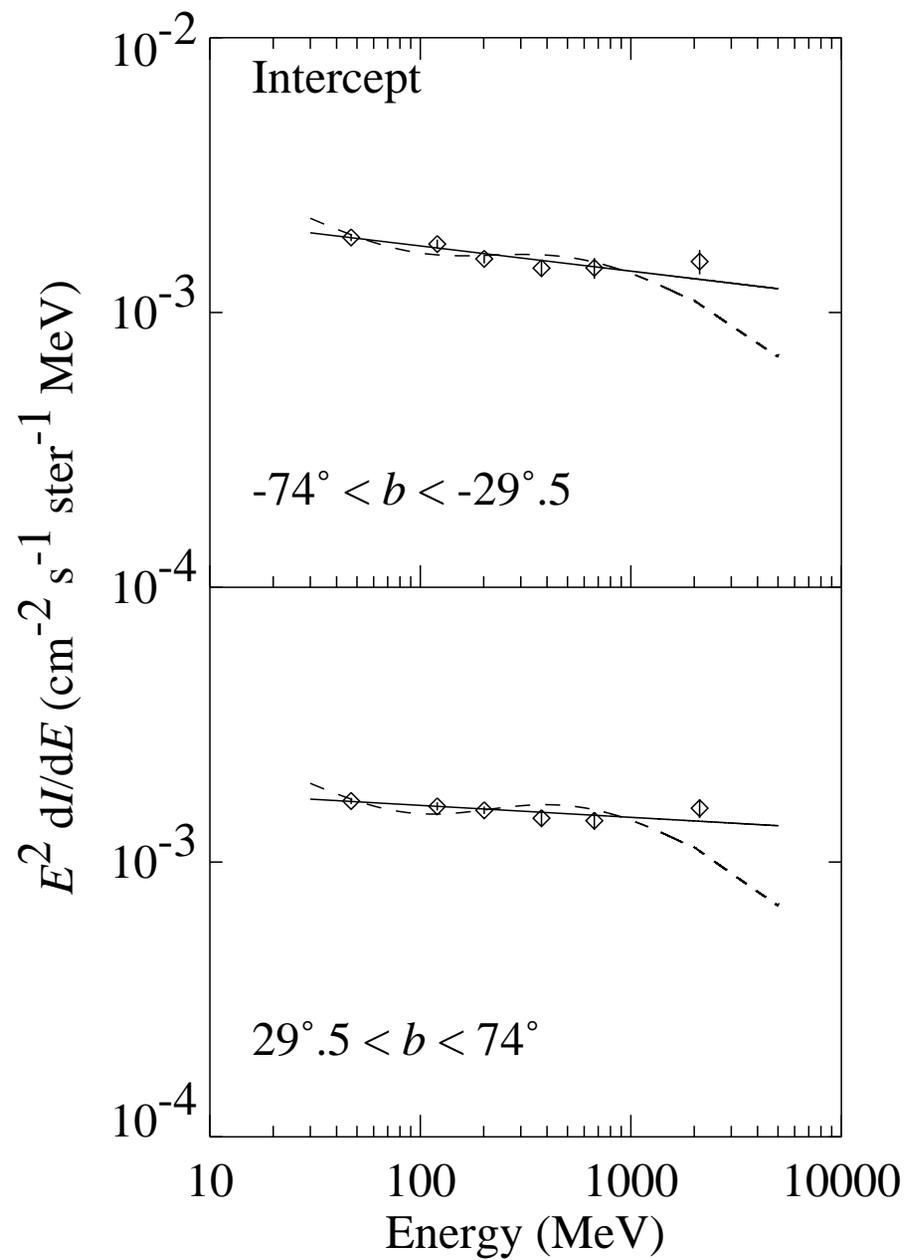

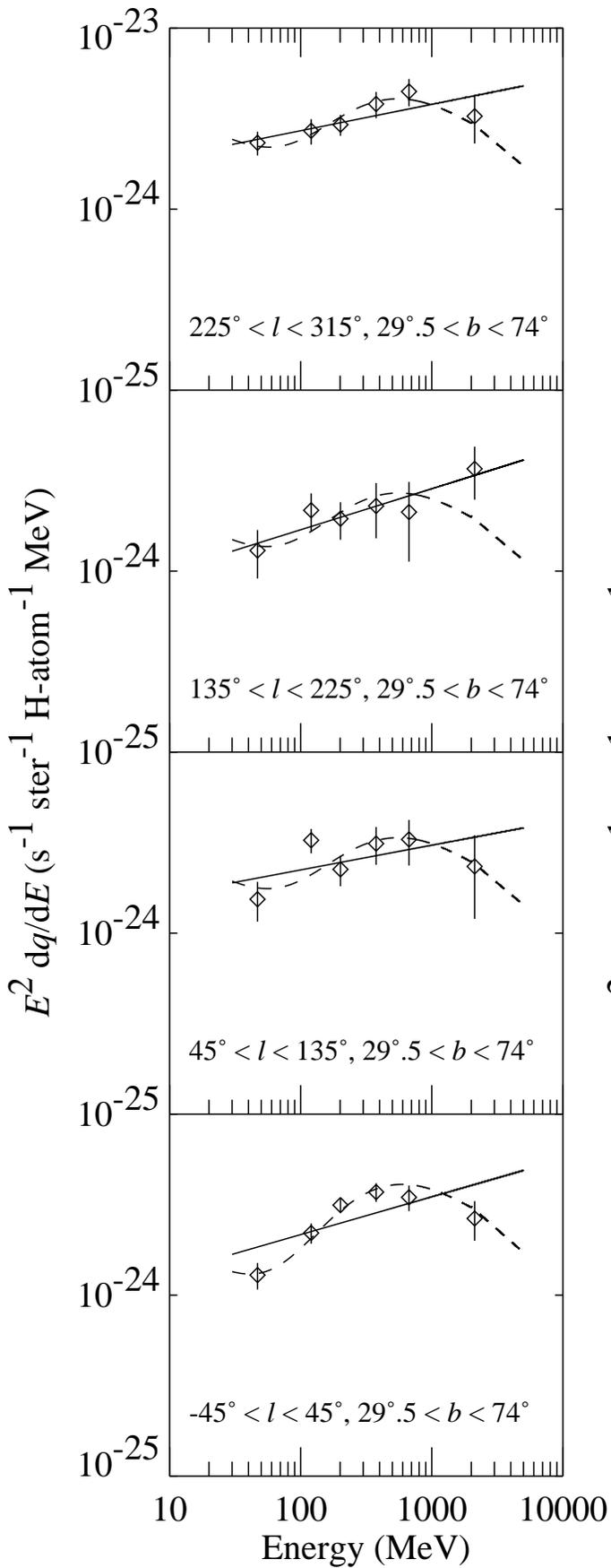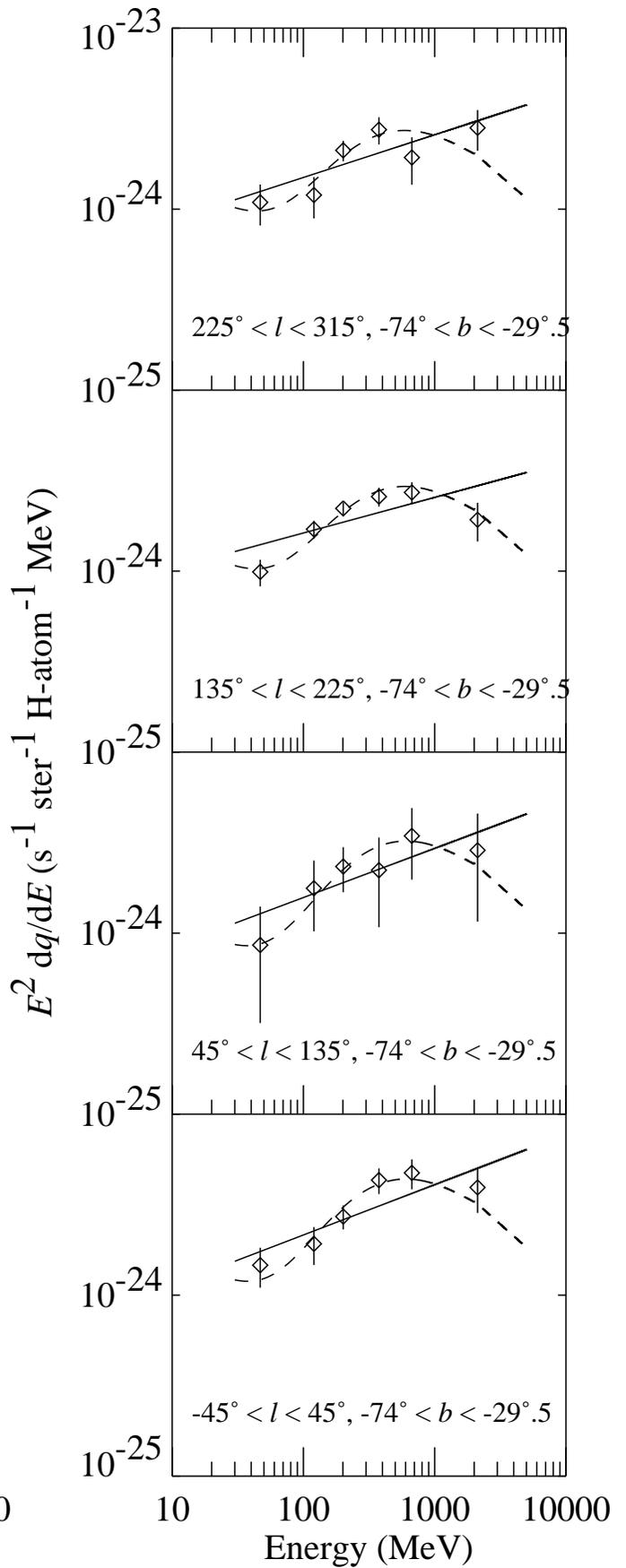

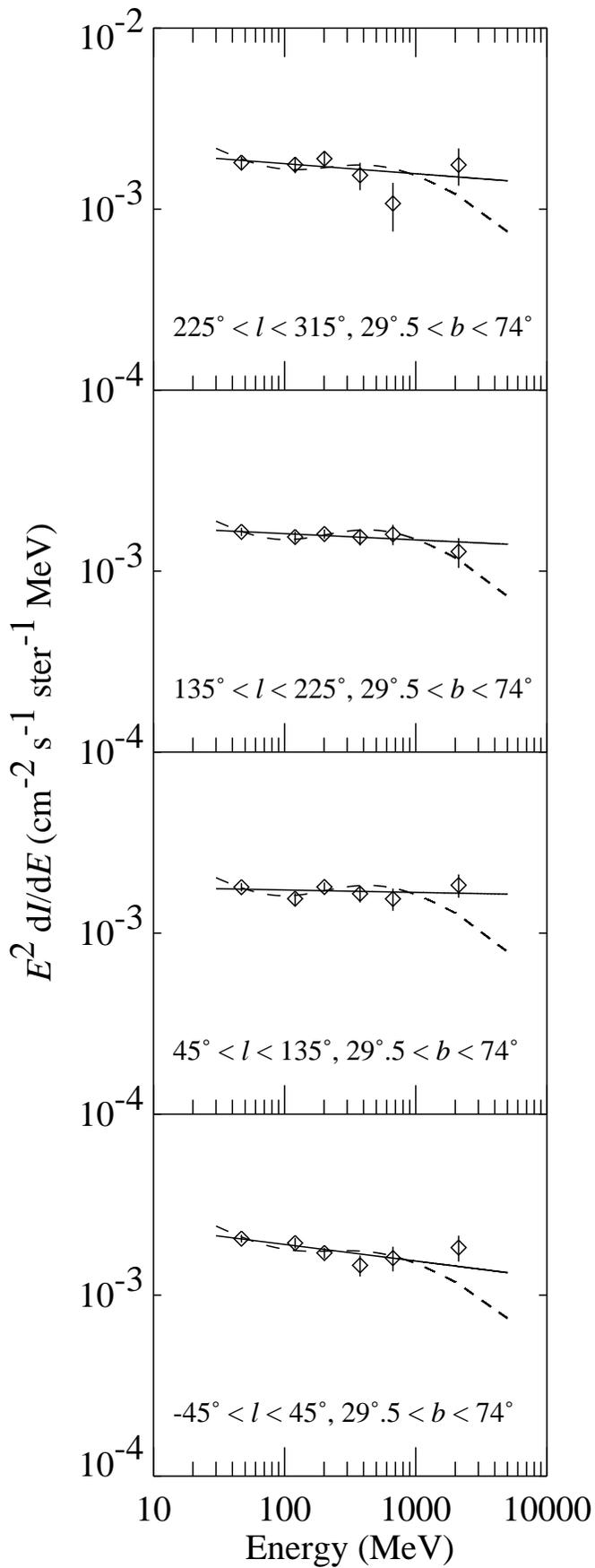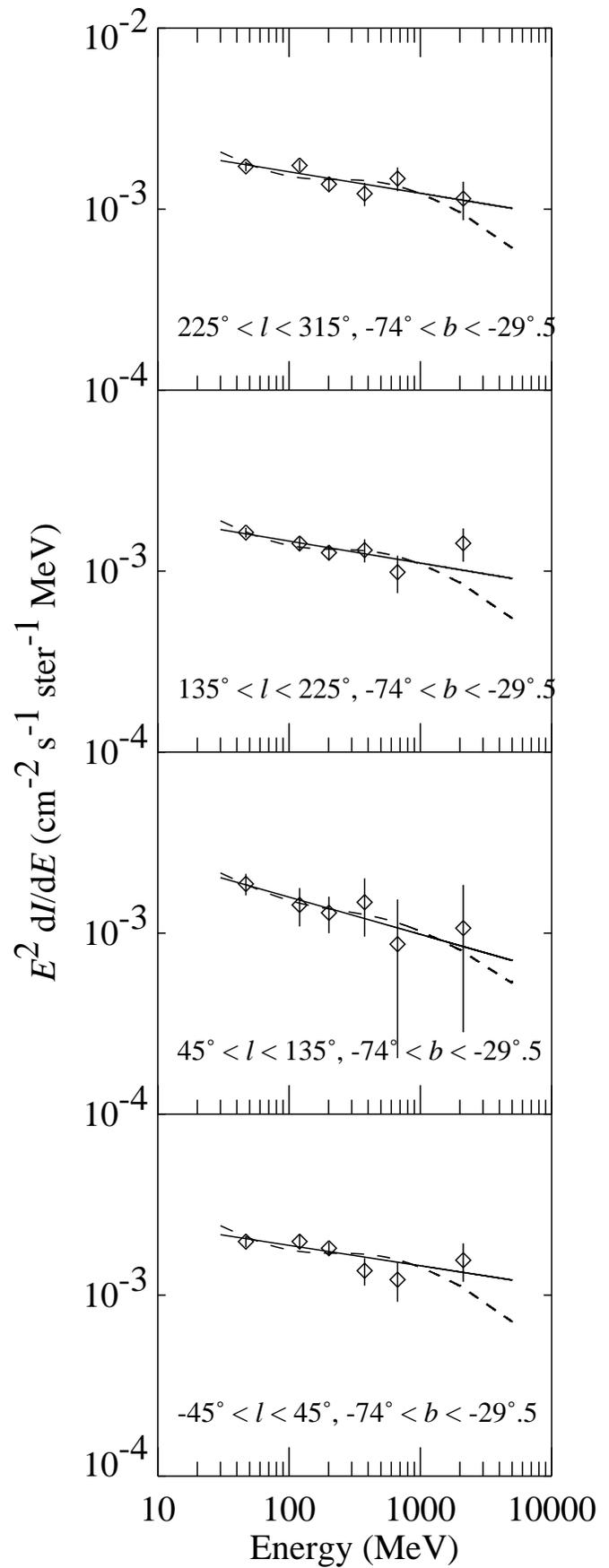

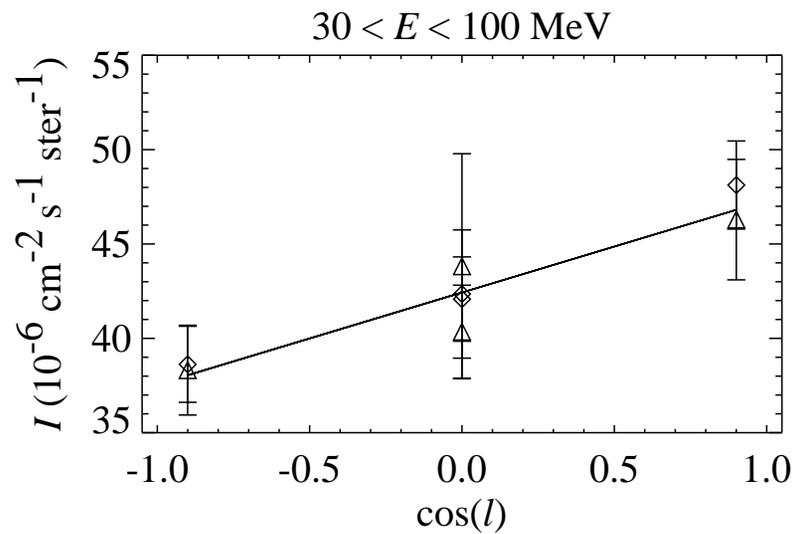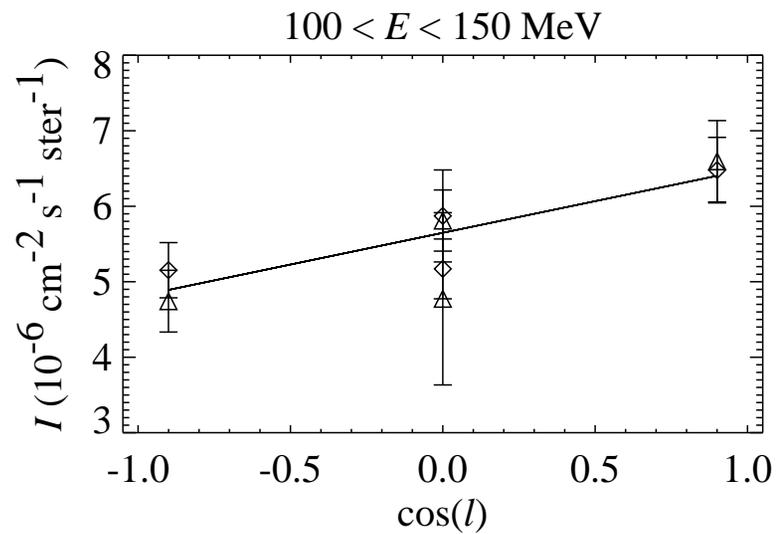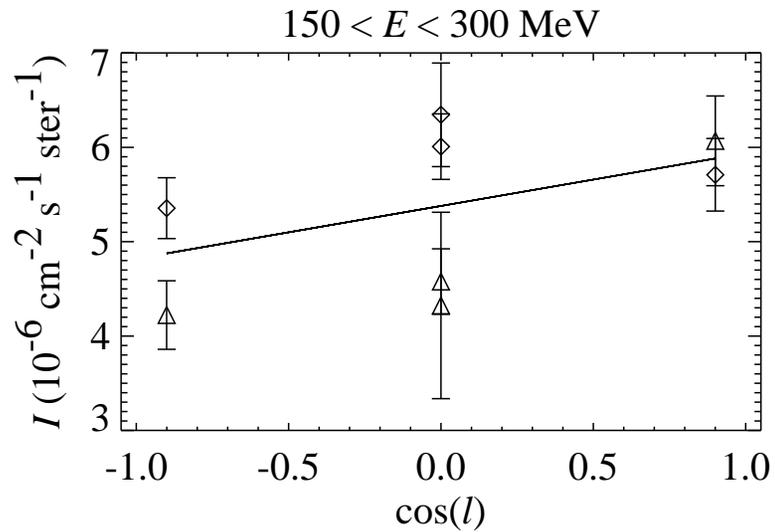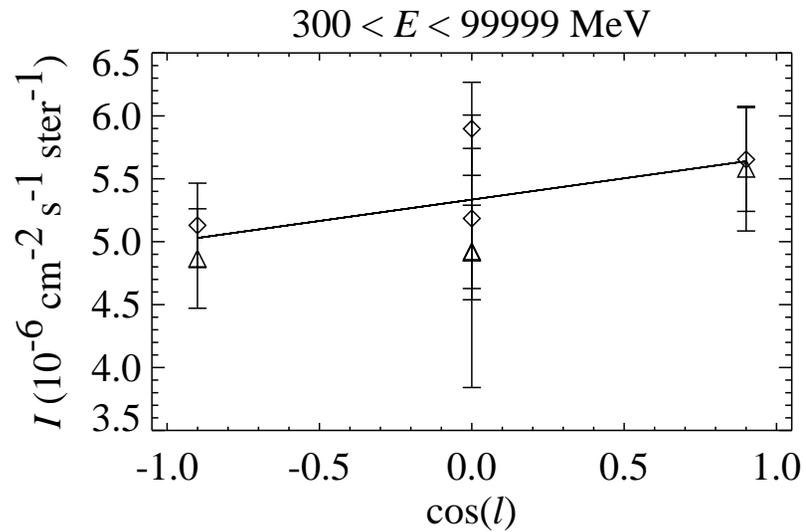

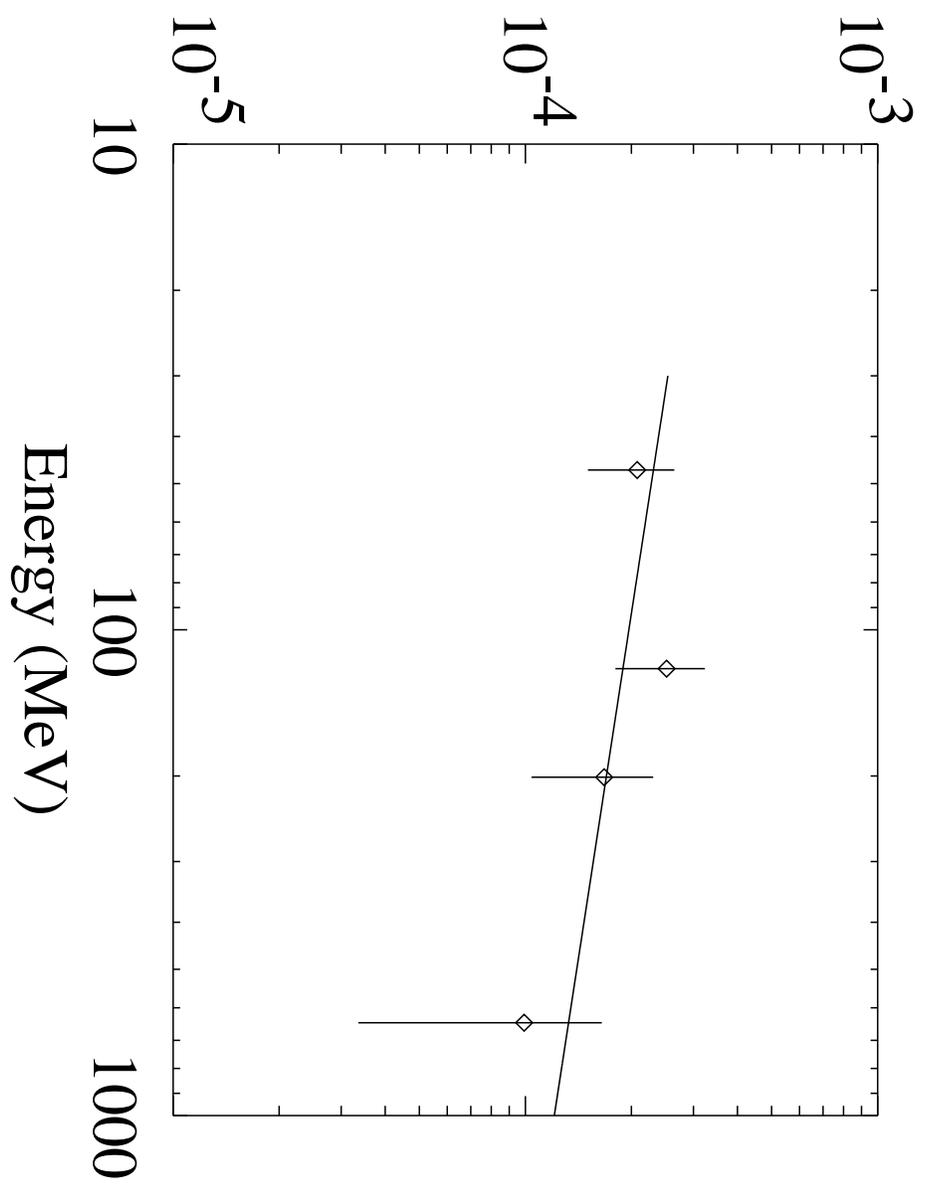

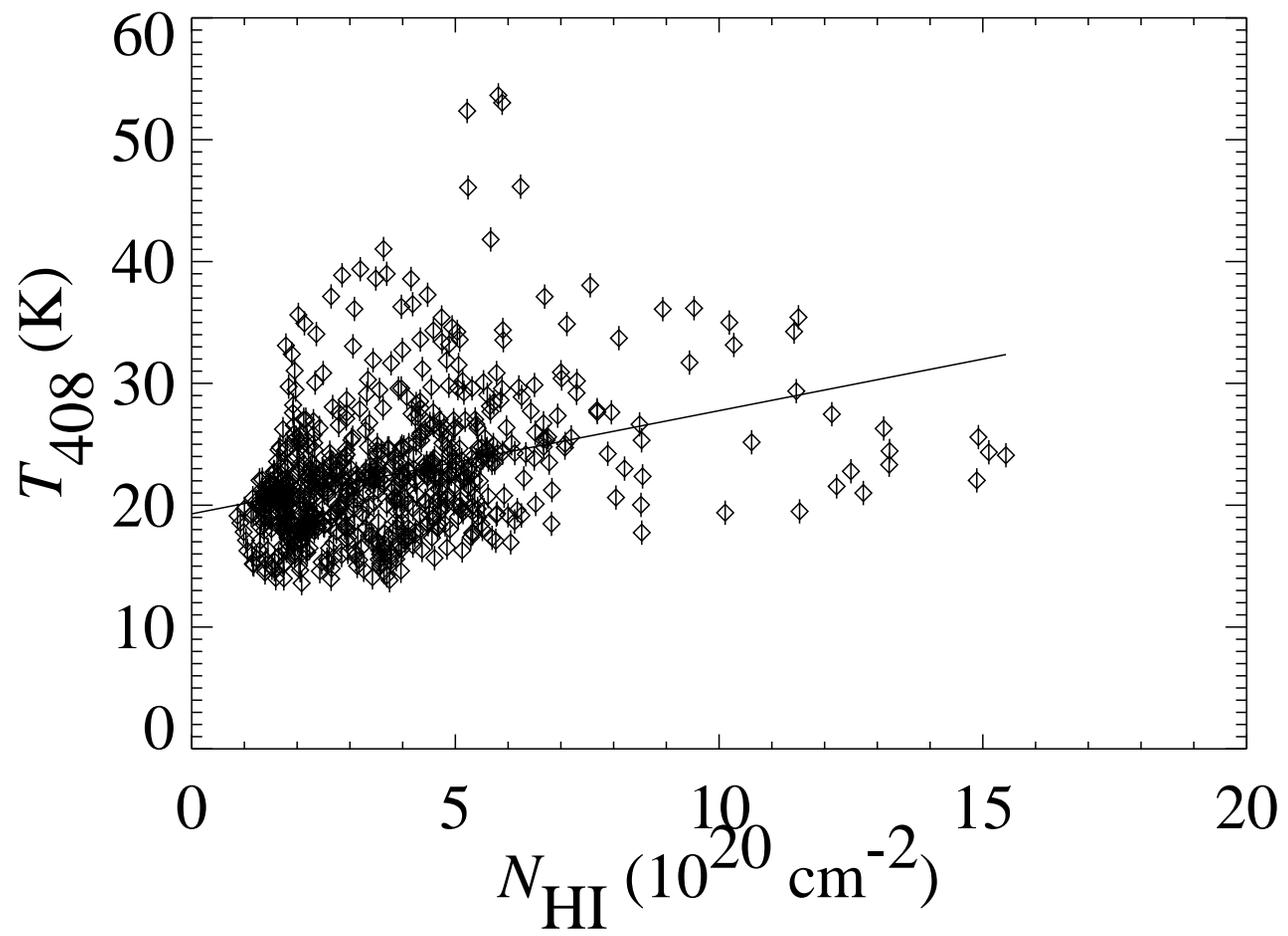

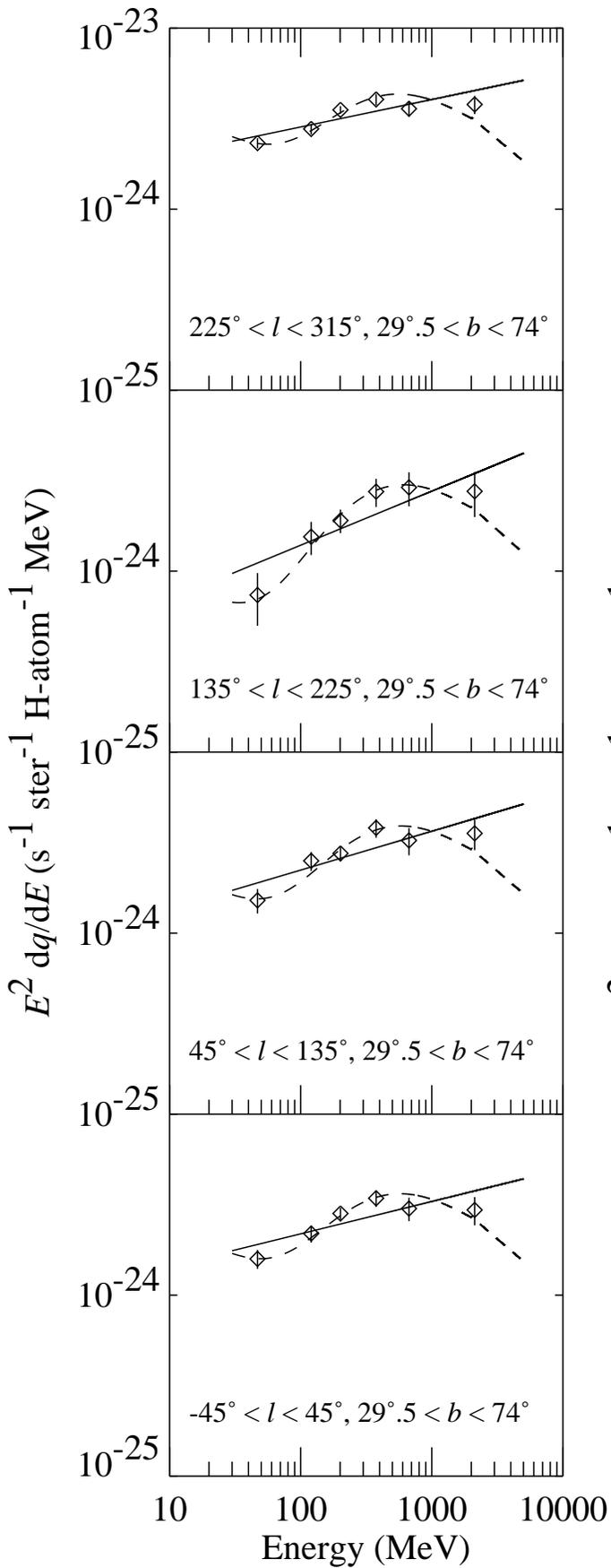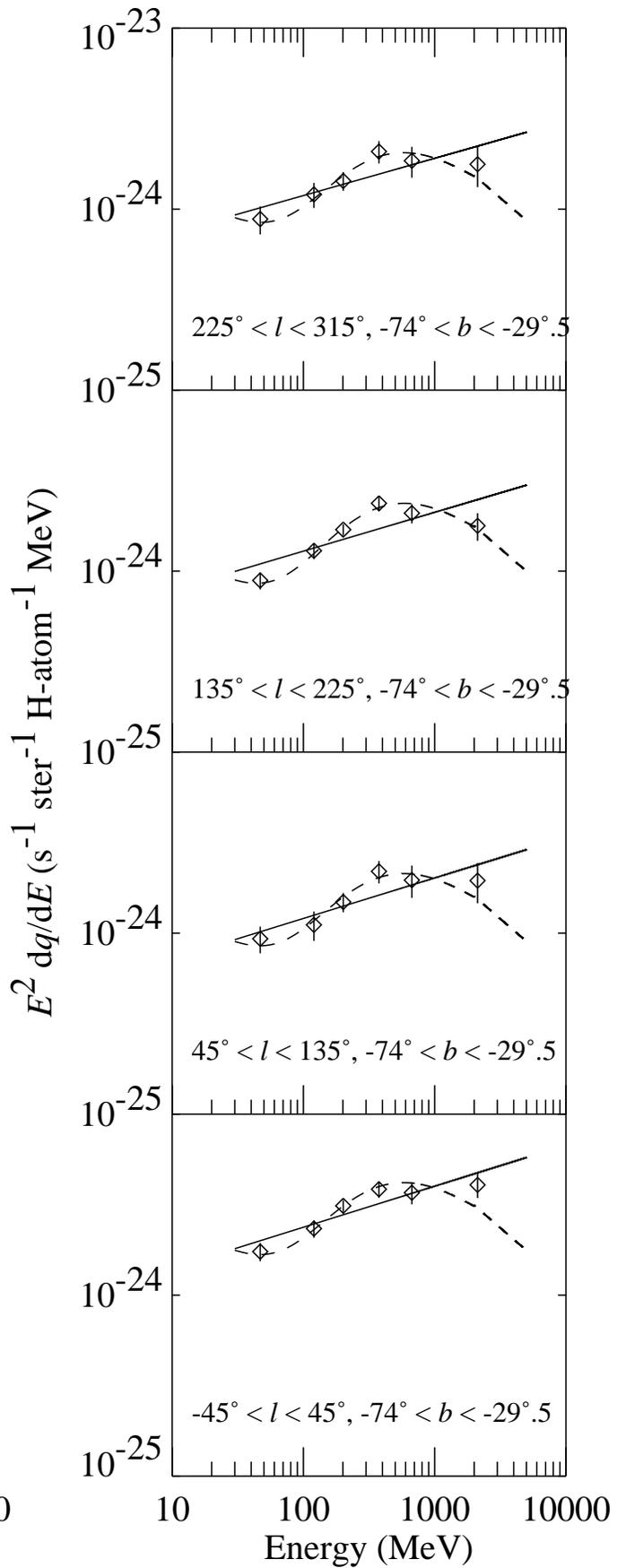

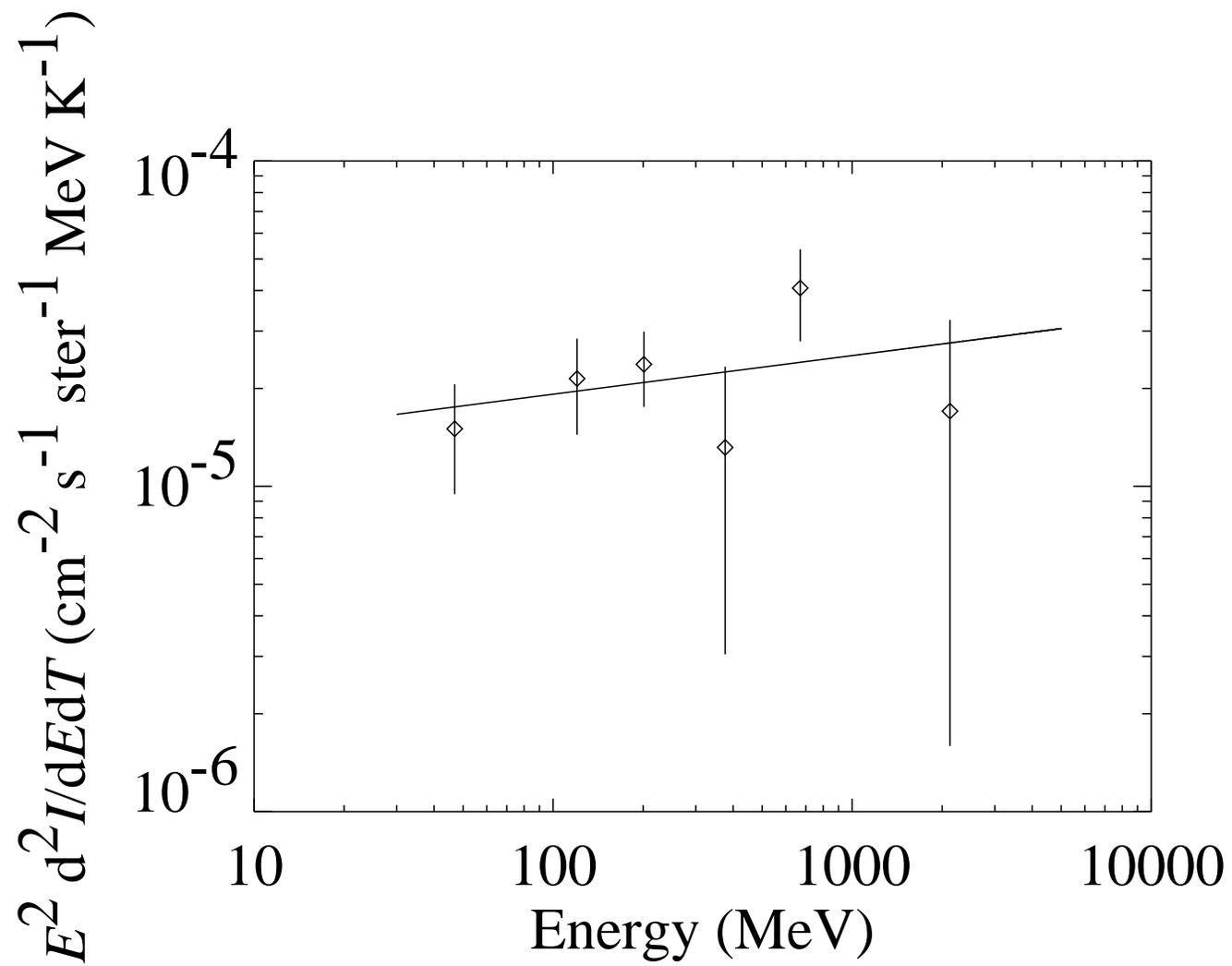

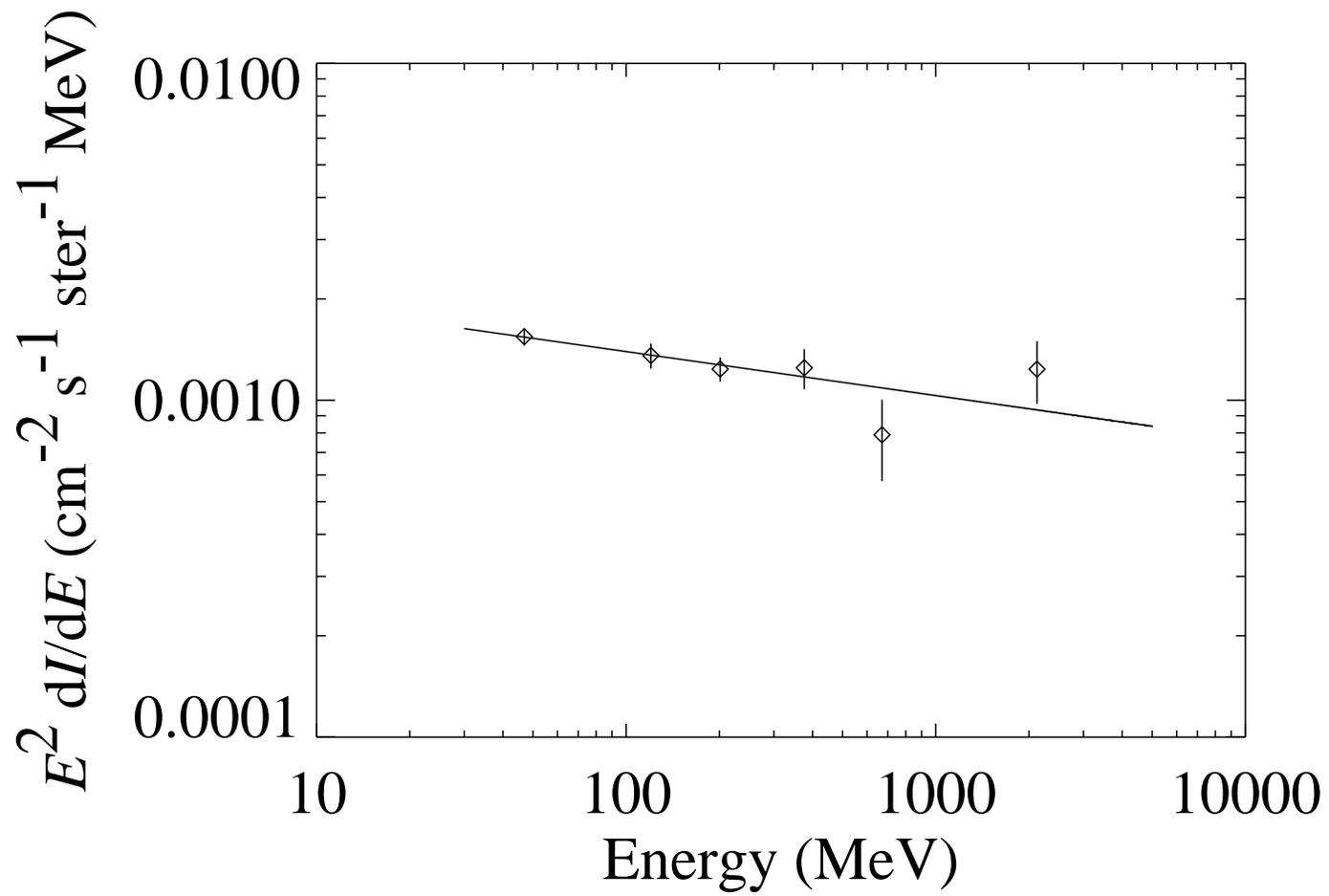